\documentclass[11pt,twoside]{article}
\usepackage{asp2006}
\usepackage{epsf}
\usepackage{lscape}
\usepackage{graphicx}
\begin{document}
\title{Mapping the Atmospheres of Hot Jupiters}
\author{H. A. Knutson}
\affil{Harvard-Smithsonian Center for Astrophysics, 60 Garden St MS 10, Cambridge, MA 02138 USA} 

\begin{abstract} 

We present the results of recent observations of phase-dependent variations in brightness designed to characterize the atmospheres of hot Jupiters.  In particular, we focus on recent observations of the transiting planet HD~189733b at 8~\micron~using the \emph{Spitzer Space Telescope}, which allow us to determine the efficiency of the day-night circulation on this planet and estimate the longitudinal positions of hot and cold regions in the atmosphere.  We discuss the implications of these observations in the context of two other successful detections of more sparsely-sampled phase variations for the non-transiting systems $\upsilon$~And~b and HD~179949b, which imply a potential diversity in the properties of the atmospheres of hot Jupiters.  Lastly, we highlight several upcoming \emph{Spitzer} observations that will extend this sample to additional wavelengths and more transiting systems in the near future.

\end{abstract}

\section{Introduction} 

There are currently more than 20 known transiting planetary systems, of which the majority are gas-giant planets orbiting extremely close ($<$0.05~A.U.) to their parent stars\footnote{See http://vo.obspm.fr/exoplanetes/encyclo/encycl.html for the latest count}.  These planets, known as ``hot Jupiters'', receive $>$10,000 times more radiation from their stars than Jupiter does from the sun, heating them to temperatures as high as 2000~K \citep{har07}.  The picture is further complicated by the fact that most of these planets are expected to be tidally locked, with permanent day and night sides.  As a result, the equilibrium compositions and circulation patterns in these atmospheres are expected to differ significantly from those of the gas-giant planets in the solar system.

It is not surprising, then, that simple models for the atmospheres of these planets \citep[see, for example, ][]{seag05,bar05,fort06b,bur07b} can sometimes differ significantly in both their assumptions and the conclusions that they reach; this is particularly true of the circulation models discussed in \S\ref{models}  Fortunately, it is possible to test the predictions of these models using observations of transiting systems.  From the wavelength-dependent depth of the transit, when the planet passes in front of the star, we can search for features in the transmission spectrum of the planet near the day-night terminator \citep{char02,vid03,vid04,knut07a,bar07,tin07,ehren07}. Similarly, by measuring the depth of the secondary eclipse, when the planet moves behind the star, we can characterize the properties of the emission from the day side of the planet \citep{char05,dem05,dem06,dem07,demory07,grill07,rich07,har07,knut07b,knut07c}.  Lastly, by measuring the changes in brightness over a significant fraction of a planet's orbit we can directly constrain the longitudinal temperature distribution across the planet's atmosphere \citep{har06,cow07,knut07b}.  This last type of observation is particularly informative for hot Jupiters, which may have extreme changes in temperature between the day and night sides of the planet.  It is also a difficult observation to make in practice as it requires a very high precision over time scales of several days; this is why all of the successful detections to date have been made from space.  Although the shape of the ingress and egress during secondary eclipse also contain information about the brightness distribution on the day side of the planet \citep{will06,rau07}, these variations are smaller and take place on shorter time scales, and there have not been any clear detections of this effect.

In this paper we focus on observations of phase variations, discussing the predictions of the current generation of atmospheric circulation models, the results of the three successful measurements of phase variations to date, and upcoming plans for more such observations in the near future.

\section{Predictions from Atmospheric Circulation Models}\label{models}

\begin{figure}
\includegraphics[angle=270]{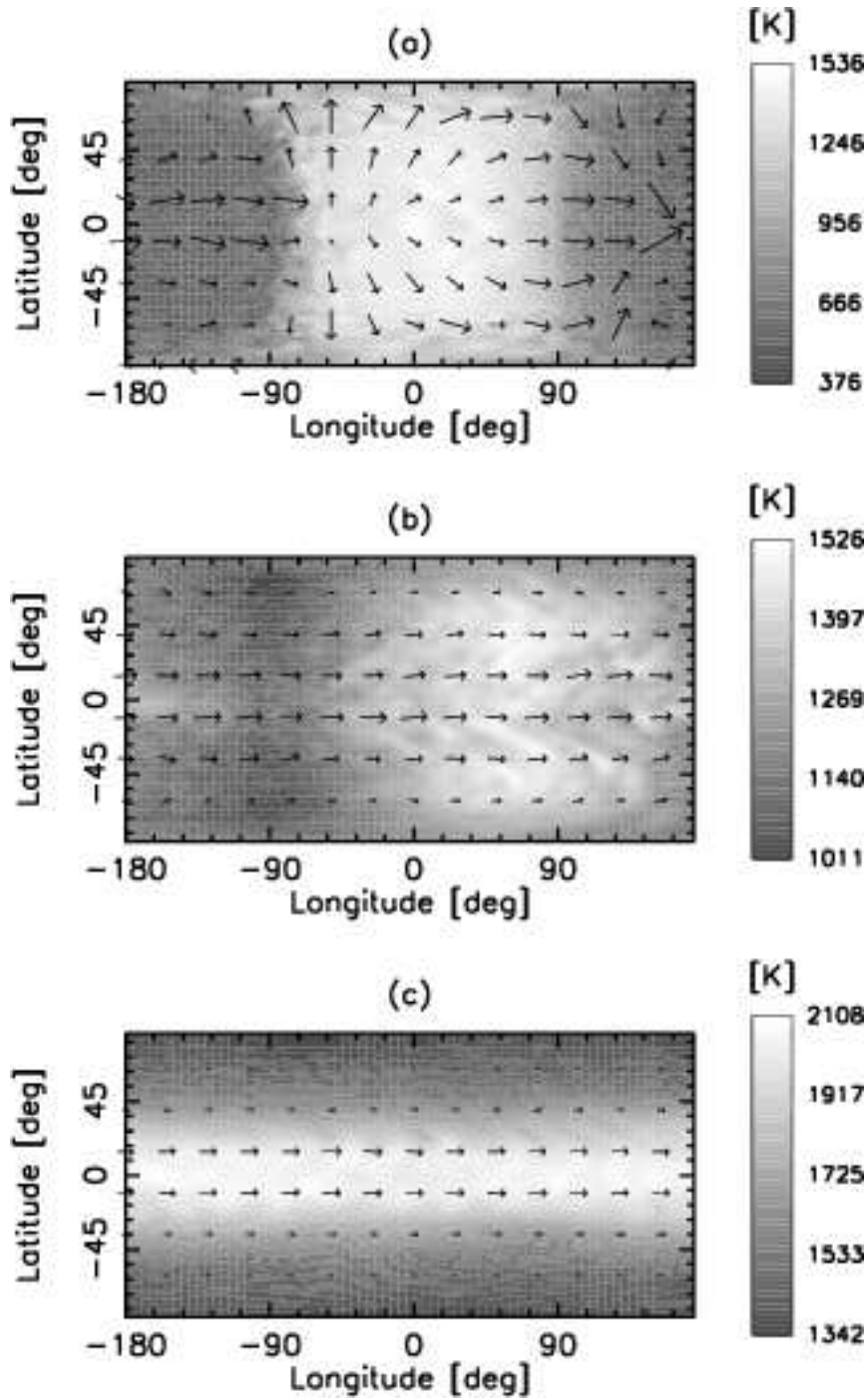}
\caption{Atmospheric circulation model for HD~209458b from \citet{coop05} showing the temperature distribution (grayscale) and winds (arrows) at three different pressures, centered on the substellar point.  Panel (a) is at 2.5 mbar, (b) is at 220 mbar, and (c) is at 19.6 bars.  This model predicts that circulation will be more efficient at depth, as demonstrated by the decreasing temperature differential and increasingly shifted position of the hot spot on the dayside in the lower panels.\label{cs_model}}
\end{figure}

As mentioned above, the extreme environments and slow rotation rates of the (presumably) tidally-locked hot Jupiters mean that their atmospheric dynamics are expected to differ significantly from those of the planets in our own solar system \citep{show02,cho03,cho06,burkert05,coop05,coop06,lang07,dobb07}.  One of the most fundamental questions is what fraction, if any, of the energy absorbed by the perpetually-illuminated day side is transferred around to the night side of the planet.  The answer to this question depends on the relative sizes of the radiative and advective time scales, quantities which are poorly constrained at the moment and may vary from planet to planet.  The relative temperature difference that we observe between the day and night sides also depends on the wavelength of the observations and the opacity of the planet's atmosphere at that wavelength.  For wavelengths where the opacity is low the effective photosphere of the planet is located deep in the atmosphere, where the pressures and temperatures are correspondingly higher.  In their dynamical models \citet{coop05} show that circulation is more efficient at these higher pressures, as the advective time scale decreases relative to the radiative time scale (see Fig. \ref{cs_model}).  Thus we would expect that observations of the same planet at different wavelengths would show varying temperature differences between the day and night sides, depending on how deep into the atmosphere we are looking at each wavelength.  

The second observationally testable prediction made by circulation models has to do with the location of hotter and cooler regions on the planet as a function of longitude.  In the most basic picture the hottest region is located at the substellar point, which has the highest levels of incident radiation, and the coolest spot is located at the antistellar point.  In atmospheres where the advective time scale is comparable to the radiative time scale the positions of these hot and cool regions are shifted in the direction of the prevailing winds.  By determine the orbital phases where the planet reaches its maximum and minimum brightness, it is possible to constrain the longitudinal positions of these hotter and colder regions.

\section{Observing Phase Variations: Two Methods}

To date there are two observing strategies that have been pursued in the quest to measure the phase variations for various planets.  In the first strategy, the same system is observed for short periods of time (usually $\sim 30$ minutes) at intervals ranging from several hours to several days.  In the second method the system is observed continuously over half an orbit, on the order of $\sim30$ hours or so.  The advantage of this second method is that it is easier to correct for instrumental effects and the resulting data has a much higher cadence and correspondingly higher precision; however it is also much more resource-intensive.

\begin{figure}
\plottwo{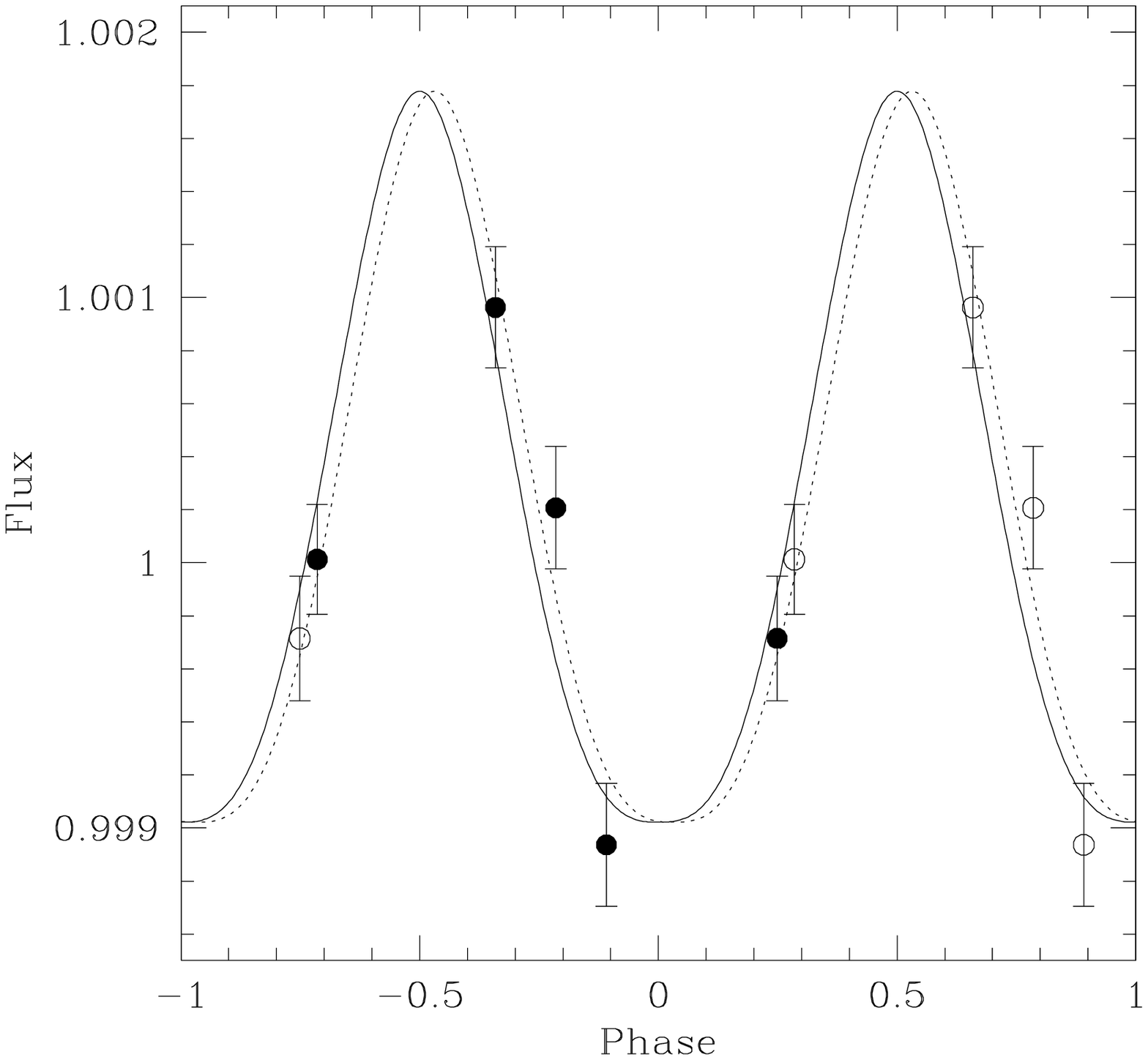}{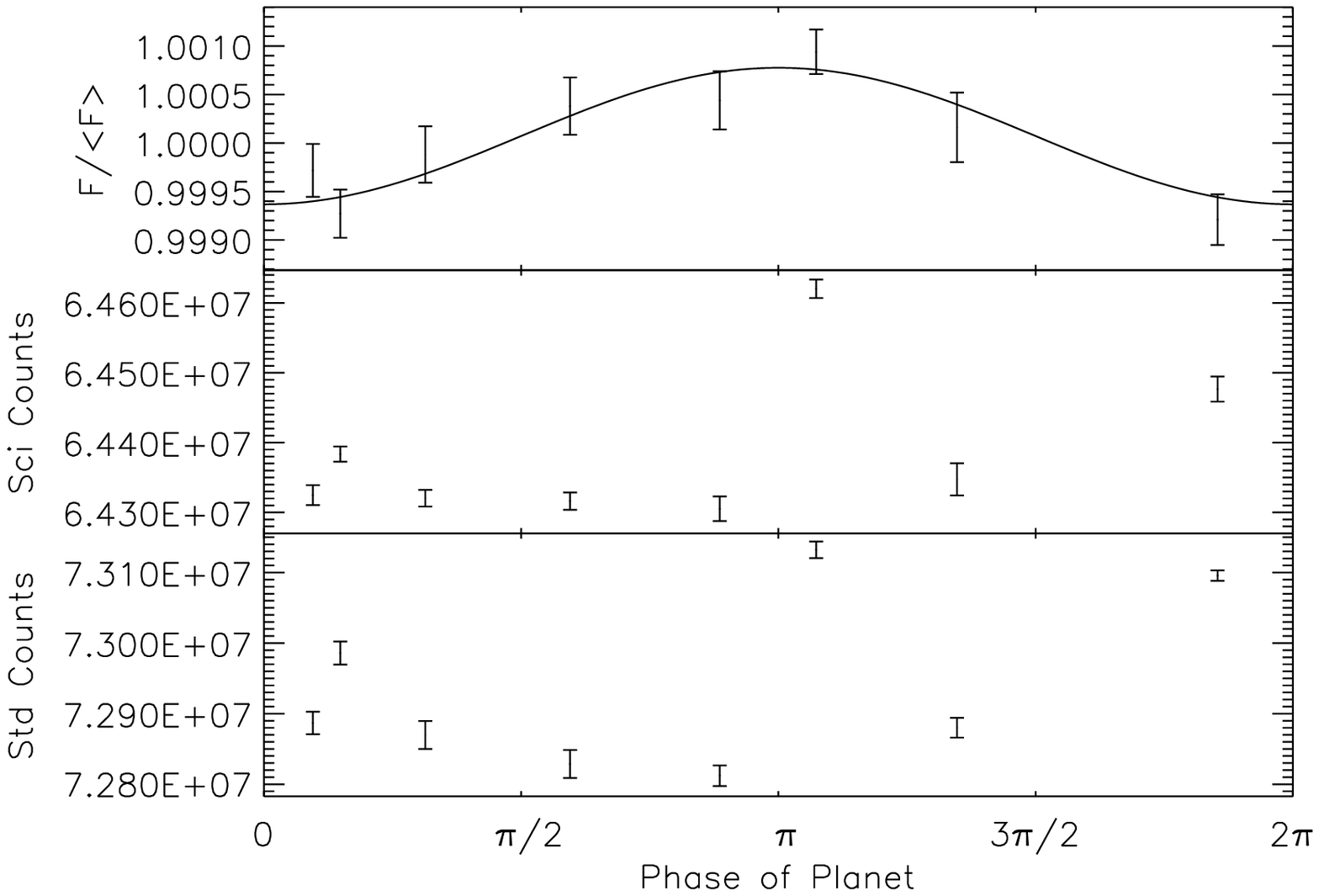}
\caption{Left panel: 24~\micron~phase variation for $\upsilon$~Andromedae b from \citet{har06}, with best fit phase curves overplotted (solid line is with no phase shift, dotted line has a phase shift).  Right panel: 8~\micron~phase curve for HD~179949 from \citet{cow07}. The two bottom panels show the 8~\micron~fluxes for HD~179949 and a comparison star used to correct for time-dependent instrument effects. The top panel shows the corrected light curve at for HD~179949 as a function of orbital phase with best-fit sinusoid overplotted. For both systems, the large amplitude of the observed variation relative to the predicted planet-star flux ratio for some reasonable set of assumptions about the planet's radius, orbital inclination, and effective temperature is consistent with relatively inefficient circulation and a large day-night temperature difference.\label{method1}}
\end{figure}

\subsection{Method 1: A Series of Widely Spaced, Short Visits}

\citet{har06} made the first detection of phase variations for an extrasolar planet by using this method to observe $\upsilon$~Andromedae b at 24~\micron~using the \emph{Spitzer Space Telescope} (see left panel of Fig. \ref{method1}).  Because this is not a transiting system we do not know the radius of the planet (although we can make an educated guess based on the mass), the inclination of the planet's orbit, or the total flux from the day side of the planet.  As a result, although \citet{har06} reported a relatively large variation in flux with orbital phase, the actual temperature difference between the day and night sides is not as well-constrained as it would be for a transiting system.  If one makes reasonable assumptions, however, the size of the observed variation implies a large day-night temperature difference and correspondingly inefficient circulation between the day and night sides.

\citet{cow07} made similar observations of three other systems at 8~\micron~using \emph{Spitzer}, including HD 209458, HD 179949, and 51 Peg, of which only one (HD 209458) is a transiting system.  They report a detection for one of the non-transiting systems, HD 179949, (see right panel of Fig. \ref{method1}) and conclude that the size of the observed variation implies a large day-night temperature difference.  They also calculate upper limits on the size of the phase variations in the other two systems, but these upper limits place relatively weak constraints on the amount of circulation in the atmospheres of the two planets.

\subsection{Method 2: Continuous Observations Over Half an Orbit}

\citet{knut07b} observe the transiting planet system HD~189733 continuously over half an orbit (33 hours) at 8~\micron~using \emph{Spitzer}.  We began our observations before the start of the transit, when the night side of the planet is visible, and end after the secondary eclipse when the day side of the planet has rotated into view.  Because we observe both the transit and secondary eclipse we are able to derive new values for the radius of the planet (from the depth of the transit) and the total flux from the day side (from the depth of the secondary eclipse), and we know we are viewing the system in an effectively edge-on configuration.  This allows us to translate the relative increase in flux over the observed part of the orbit into a longitudinal temperature distribution across the surface of the planet, as described below.  We note that the presence of spots on the star, which is known to be quite active \citep{winn07,henry07}, and corrections for detector effects \citep{char05,mor06} both contribute additional uncertainties in the interpretation of this measurement; both effects are discussed in more detail in \citet{knut07b}.

\begin{figure}
\plottwo{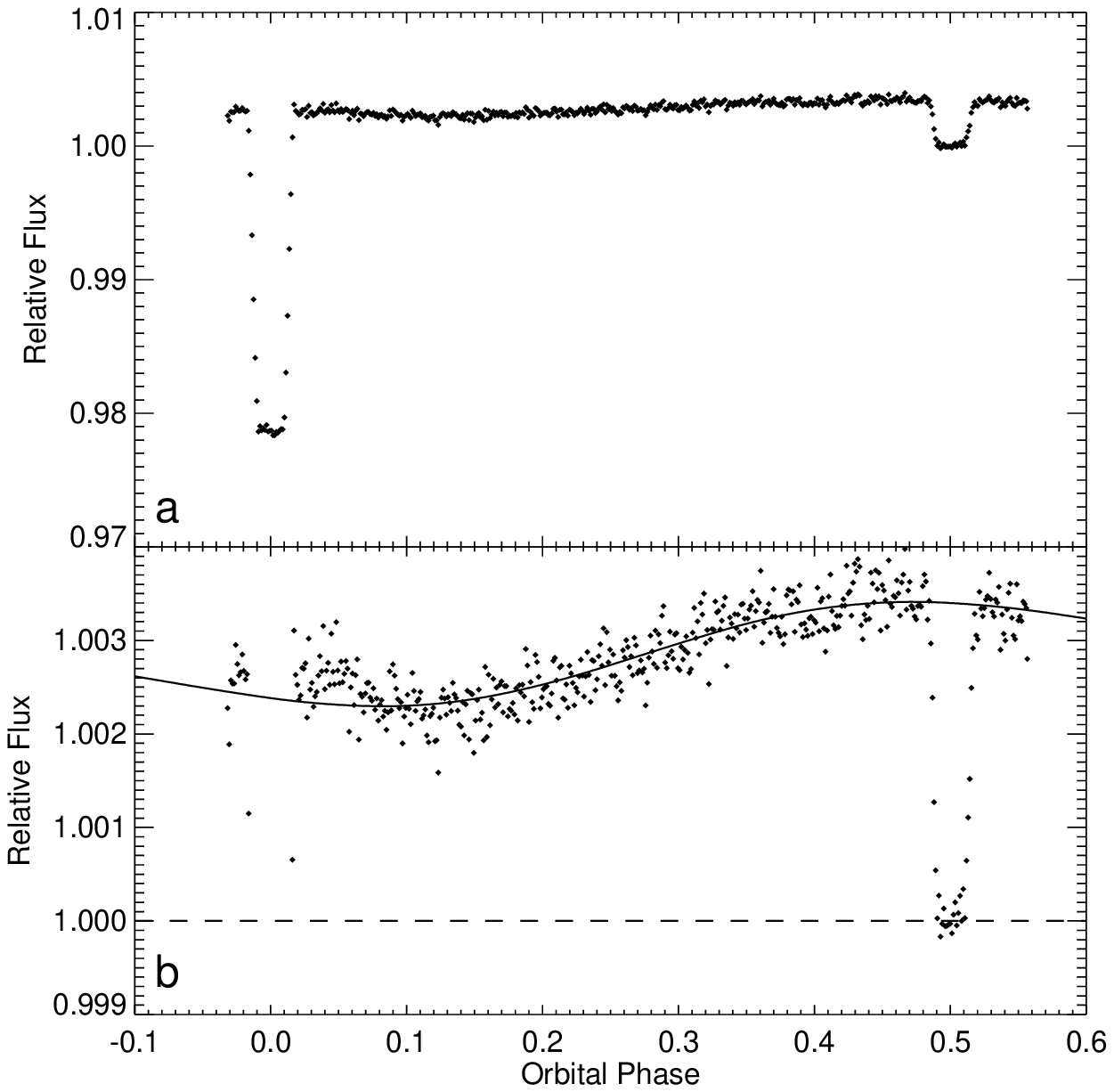}{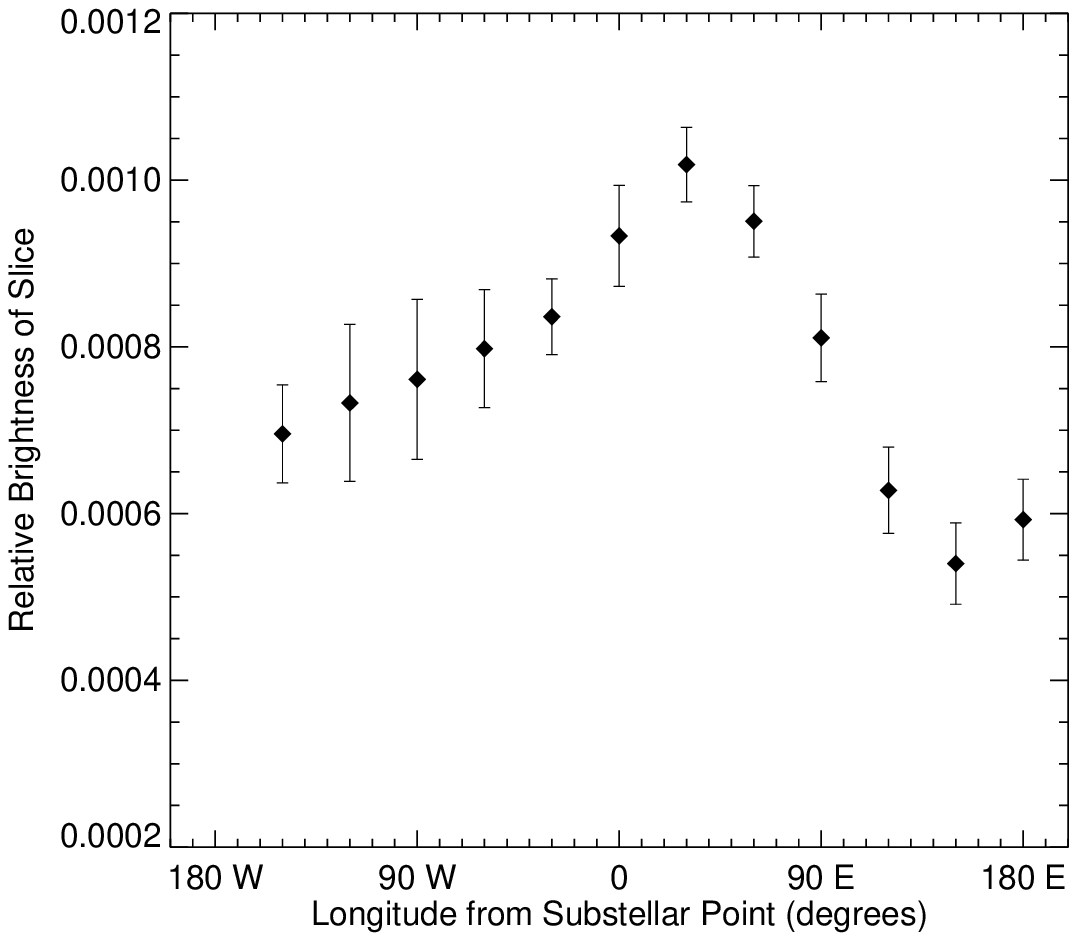}
\caption{Left panel: Observed phase variation for HD 189733b from \citet{knut07b}, with transit and secondary eclipse visible.  The stellar flux as measured at the center of the secondary eclipse is normalized to unity (dashed line), and the data is binned every 500 points (200 s).  {\bf a} and {\bf b} show the same data, but in {\bf b} the y axis is expanded to show the scale of the variation.  The solid line is the phase curve for the best-fit model shown on the right.  Right panel: Brightness estimates for twelve longitudinal strips on the surface of HD 189733b.  The brightness estimates are given as the ratio of the flux from an individual slice viewed face-on to the total flux of the star.\label{method2}}
\end{figure}

We measure a total increase in flux of $0.12\pm0.02\%$ over the period of our observations (see Fig. \ref{method2}).  If we scale this value by the depth of the secondary eclipse, this means that the minimum hemisphere-integrated flux is $64\pm7\%$ of the maximum hemisphere-integrated flux.  Converting this to brightness temperatures, we find that the minimum hemisphere-averaged brightness temperature is $973\pm33$~K, and the maximum hemisphere-averaged brightness temperature is $1212\pm11$~K.  The uncertainties in the minimum brightness temperature, which approximately corresponds to the night side temperature, are larger because this part of the planet was visible at the beginning of the observations when the uncertainties in the correction for the detector ramp are largest.

We also find that the minimum brightness occurs $6.7\pm0.4$ hours after the center of the transit, while the maximum brightness occurs $2.3\pm0.8$ hours before the center of the secondary eclipse.  These shifts imply that the hottest and coolest points are not located at the substellar and antistellar points, but are instead shifted to the east and to the west, respectively.  We can refine our estimates of the locations of these spots by fitting the observed phase variation with a model made of twelve longitudinal strips of varying brightness (Fig. \ref{method2}); the resulting brightness distribution implies that the hot spot is shifted approximately 30 degrees east of the substellar point, while the cool spot is shifted 30 degrees west of the antistellar point.  The shifted hot spot is consistent with the west-to-east winds predicted by most models \citep{show02,cho03,cho06,burkert05,coop05,coop06,lang07,dobb07}, but these models also predict that the cool spot should also be shifted to the east.

\subsubsection{Diversity of Planetary Atmospheres}

There are several conclusions we can draw from these data.  First, unlike $\upsilon$~And~b and HD~179949b, we find that for HD~189733 the circulation between the day and night sides of the planet must be relatively efficient in order to produce such a small temperature difference between the day and night sides.  Although it is possible that some of the difference between $\upsilon$~And~b and HD 189733b may be due to the fact that the $\upsilon$~And~b observations were at a longer wavelength (24~\micron~instead of 8~\micron), it is unlikely that this difference alone would produce such a large change in the amplitude of the observed phase variation.  Furthermore, this cannot account for HD 179949b's large phase variation, as this system was observed at the same wavelength as HD 189733b.  

In light of the recent detection of a atmospheric temperature inversion on the dayside of HD~209458b by \citet{knut07c}, it is more plausible to think that the atmosphere of HD 189733b has fundamentally different properties than those of $\upsilon$~And~b and HD 179949b.  There are some initial indications that the presence of a temperature inversion may be correlated with the level of irradiation and/or the surface gravity of the planet in question \citep{bur07a,bur07b,fort07}, as evidenced by the fact that TrES-1 and HD~189733b do not appear to have temperature inversions \citep{char05,dem06,rich07}, while HD 209458b and HD 149026b probably do \citep{grill07,knut07b,bur07a,fort06a,har07}.  $\upsilon$~And~b and HD 179949b both have higher levels of incident flux than HD 209458b; thus it would not be unreasonable to think that both of these planets may have temperature inversions. Such inversions would naturally lead to a larger phase variation, as the incident flux is absorbed higher in the atmosphere \citep{fort07}.

\section{Conclusions and Future Directions}

The above discussion highlights the significant advances that have been made in the year since \citet{har06} reported the first detection of the phase variation of an extrasolar planet.  However, the current sample is limited, and there is clearly much to be gained from observations of additional transiting systems, for which the planetary radius, orbital inclination, and dayside flux are all well-constrained.  There are a number of exciting programs scheduled for the upcoming \emph{Spitzer} observing cycle which will address this issue.  First, we have a program which will observe HD 189733b continuously at 24~\micron~over the same half orbit as \citet{knut07b}, which should probe a different region of the atmosphere than the previous 8~\micron~observations.  As a part of this same program we will also observe HD~209458b over half an orbit at both 8 and 24~\micron, which will allow us to directly compare the wavelength-dependent properties of the atmospheric circulation for these two planets. 

To date all of the observations we have described have examined planets with effectively circular orbits; however, there is another promising set of upcoming observations by G. Laughlin and G. Bakos that will observe two highly eccentric planets (HD 80606b and HAT-P-2b) continuously over $\sim30$ hours centered on periastron passage.  Although only one of these planets, HAT-P-2b, is transiting, both observations should detect the rapid increase in brightness as the planetary atmosphere is flash-heated during this passage, providing a direct measurement of the radiative time constant.  

Of the two methods described above, it is the more time-intensive, continuous observations of transiting systems that have provided the most detailed information about circulation in the atmospheres of hot Jupiters to date.  However, the detections of phase variations for $\upsilon$~And~b and HD~179949b highlight the advantages of more sparsely-sampled observations in facilitating a much wider-ranging survey of a large number of planetary systems.  These two approaches are inherently complementary, and we expect that both will make valuable contributions to our understanding of these unusual planets in the near future.

\acknowledgements 

This work is based on observations made with the \emph{Spitzer Space Telescope}, which is operated by the Jet Propulsion Laboratory, California Institude of Technology, under contract to NASA.  Support for this work was provided by NASA through an award issued by JPL/Caltech.  HAK was supported by a National Science Foundation Graduate Research Fellowship.

\end{document}